# Odd integer quantum Hall states with interlayer coherence in twisted bilayer graphene

Youngwook Kim[1,2], Pilkyung Moon[3], Kenji Watanabe[4], Takashi Taniguchi[5], and Jurgen H. Smet[1*]

[1]*Max-Planck-Institut für Festköperforschung, 70569 Stuttgart, Germany*

[2]*Department of Emerging Materials Science, DGIST, 42988 Daegu, Korea*

[3]*Arts and Sciences, NYU Shanghai, Shanghai 200122, China and NYU-ECNU Institute of Physics at NYU Shanghai, Shanghai 200062, China*

[4]*Research Center for Functional Materials, National Institute for Materials Science, Tsukuba 305-0044, Japan*

[5]*International Center for Materials Nanoarchitectonics, National Institute for Materials Science, Tsukuba 305-0044, Japan*

**Abstract:**

**We report on the quantum Hall effect in two stacked graphene layers rotated by 2°. The tunneling strength among the layers can be varied from very weak to strong via the mechanism of magnetic breakdown when tuning the density. Odd-integer quantum Hall physics is not anticipated in the regime of suppressed tunneling for balanced layer densities, yet it is observed. We interpret this as a signature of Coulomb interaction induced interlayer coherence and Bose Einstein condensation of excitons that form at half filling of each layer. A density imbalance gives rise to reentrant behavior due to a phase transition from the interlayer coherent state to incompressible behavior caused by simultaneous condensation of both layers in different quantum Hall states. With increasing overall density, magnetic breakdown gains the upper hand. As a consequence of the enhanced interlayer tunneling, the interlayer coherent state and the phase transition vanish.**

The study of two-dimensional electron systems in close proximity has been a very rewarding subject in the III-V semiconductor community. The observation of Bose Einstein condensation is an example of one of the most prominent outcomes.[1] The rich physics in these coupled electron systems stems from the additional degrees of freedom they offer. Tunneling allows for the transition of charge carriers between the layers and interlayer Coulomb interactions compete with intralayer interactions. Both can be tuned by changing the characteristics of the tunnel barrier and the interlayer separation. For small layer spacing, interlayer tunneling mixes the levels of each layer, forming symmetric and antisymmetric states. In III-V semiconductors, the resulting energy gap is usually smaller than the cyclotron energy at the typical magnetic fields, but comparable with the interlayer Coulomb interaction energy. This often leads to incompressible ground states distinct from those found in single layer systems. It causes a magnetic field driven collapse of the tunneling induced energy gap as well as the formation of a Bose Einstein condensate for instance at half filling of each layer, heralded by quantum Hall behavior at odd integer filling.[2-7]

The state-of-the-art mobilities offered by GaAs double layer systems were instrumental for the discovery of this condensate as well as Josephson like tunneling behavior and vanishing Hall resistance in counter-flow experiments, considered a signature of excitonic superfluidity[3-4]. More recently, the flexibility in device design and sample quality offered by graphene layers separated by an insulator with atomic flatness and thickness control through van der Waals stacking and encapsulation has caused the community to revisit this field in order to benefit from the exceptional charge carrier tunability offered and the expected enrichment of the physics due to the additional valley degree of freedom. Analogous phenomena such as those observed in GaAs were first reported, including the quantum Hall state at total filling 1 due to Bose Einstein condensation, as well as vanishing Hall resistance in counter-flow experiments.[8,9] However, also more exotic interlayer paired states, like the $\nu_{tot} = 1/3$ state, have been discovered.[10,11]

Here, we address the extreme regime of sub-nanometer layer spacing and yet strongly suppressed interlayer coupling or tunneling, that has not been at the focal point so far. To this end, we exploit the virtues of twisted graphene layers without an insulating layer in between. As a result of the twist, the Dirac cones from each layer are displaced in momentum space as illustrated in Fig. 1a, which can strongly suppress the interlayer coupling or tunneling.[12] Near the so-called magic angle of 1.05°, the band dispersion becomes flat and both layers are always strongly coupled. This gives rise to a rich phase diagram with ferromagnetic, correlated insulating and superconducting ground states.[13-19] Of interest here are larger twist angles that produce an energy dispersion in the reduced Brillouin zone closely resembling that of graphene with a van Hove singularity due to the merger of the displaced Dirac cones of the original layers (see Fig. 1a). The singularity occurs at an energy no longer set by the graphene hopping parameter (about 3eV), but a much lower twist angle controlled energy. Prior to reaching the van Hove singularity, magnetic breakdown due to the uncertainty in reciprocal space in a magnetic field[20-22] allows tunneling across momentum gaps between the disjoint and closed Fermi surfaces, that form in the reduced Brillouin zone (Fig. 1b). The original Dirac cone states at the displaced $K$-symmetry points give rise to a band with closed Fermi surfaces encircling the $\bar{K}$ and $\bar{K}'$ points of the reduced Brillouin zone (solid circles). The coloring refers to the layer the states originate from (black for layer 1 and blue for layer 2). The same holds for the Dirac cone states at the displaced $K'$-symmetry points, however with an interchanged location in the reduced zone (dotted Fermi surfaces). The momentum gaps at the reduced zone boundary shrink with increasing density (Fig. 1b) and the tunneling probability rises. It is this mechanism that accounts for enhanced "hybridization" or "coupling" between the different layer states with increasing density. The system can then no longer be regarded as composed of two separate layers. When the chemical potential crosses the van Hove singularity, the Fermi surfaces of each band merge into a single Fermi surface effectively removing the layer degree of freedom all together.

At small angles above the magic angle, the chemical potential can easily be lifted up to this energy with conventional gating techniques.[23,24] Hence, it is possible to explore the regime of essentially decoupled layers at low densities, weakly tunnel coupled layers at intermediate densities and strongly coupled layers when the chemical potential comes closer to the van Hove singularity. All this can be accomplished at the turn of a gate voltage knob. Devices with twist angles between 2° to 3° represent the sweet spot, because the Fermi velocity in the low energy regime is significantly reduced compared to devices with larger twist angles. The enhanced density of states boosts the importance of Coulomb interactions. Moreover for twist angles exceeding 5° the overlap matrix element[25] or interlayer transitions between eigenstates of the opposite valleys $K$ and $K'$, i.e.

between the two Fermi surfaces encircling either the $\bar{K}$ and $\bar{K}'$ points of the reduced Brillouin zone (Fig. 1b) grows rapidly, disrupting the separate layer picture.

As opposed to previous studies focusing either on large twist angles[26-28] or the magic angle,[13-19] here we address the transport properties of a device with a twist angle of 2° to search for Bose Einstein condensation related quantum Hall states for the above mentioned reasons. For this angle, the van Hove singularity energy is only about 25 meV.[23,24] A device schematic and image of the are shown in panels a and b of Fig. 2. A bottom and top gate enable the application of a displacement electric field $D$. Further details of the device and gating characteristics are deferred to the supplementary information. A color map of the longitudinal resistance recorded across the plane spanned by the total density $n_{tot}$ and $D/\varepsilon_0$ is plotted in Fig. 2c. Here, $\varepsilon_0$ is the permittivity of vacuum. Key features are highlighted: the charge neutrality point at $n_{tot} = 0$ and the two satellite peaks signaling full occupation or depletion of the moiré potential induced miniband centered around zero energy. The latter appear at a hole and electron density of approximately $8.5 \times 10^{12}$ cm$^{-2}$ (section S2, supporting information). The density accommodated by the miniband yields the area of the reduced Brillouin zone from which it is possible to determine the twist angle of 2°. The same twist angle can also be obtained from the Landau fan chart.

In Figure. 2c, the charge neutrality peak at the Dirac point rapidly diminishes in amplitude (from red to yellow) when applying an electric field across the layers, while the strength of the two satellite peaks remains nearly constant. This behavior reflects that when the chemical potential is near zero energy, the two layers are decoupled due to the momentum mismatch of the displaced $K$ and $K´$ symmetry points for the two layers. The displacement field does not cause a band gap as in Bernal stacked bilayer graphene, but shifts the chemical potential with respect to the Dirac zero energy points of the two graphene sheets in opposite direction. There is then a non-zero density of states in both graphene layers at the chemical potential. Holes accumulate in one layer and electrons in the other layer resulting in low resistivity, while the net density remains zero. In contrast, near the secondary charge neutrality points the resistivity is insensitive to the displacement field. This comes as no surprise, since beyond the van Hove singularity, the layers are strongly coupled and the resistance peak results from the twist angle induced bandgap separating adjacent minibands. This behavior due to miniband formation is consistently observed both in magic angle devices as well as in a 2° twisted device.[12-19, 23-25,29]

In Figure 3a the longitudinal magnetoresistance and the Hall conductance are plotted when the sample is exposed to a fixed perpendicular magnetic field of 3T. At zero displacement field and when the two layers are decoupled, it is anticipated that the sequence of observable QH states is governed by the 8-fold degeneracy due to the spin, valley and layer degrees of freedom at total filling $\nu_{tot}$ = …, -20, -12, -4, +4, +12, +20, …. This is confirmed in the experimental trace of Fig. 3a. A color rendition of the longitudinal resistance across the plane spanned by the displacement electric field and either the total density or total filling is shown in panel b of Fig. 3. At larger densities, both layers hybridize due to magnetic breakdown as the momentum gaps between the Fermi surface orbits shrink with increasing carrier density. The layer degree of freedom is effectively removed when the chemical potential approaches the van Hove singularity due to the Lifshitz transition from two to one Fermi surface per band. The 8-fold degeneracy is replaced by a 4-fold degeneracy. Experimental manifestations of this can be seen in Fig. 3a. The longitudinal resistivity develops in the absence of a displacement field additional minima at $\nu_{tot} = \pm 24$ and $\pm 32$, marked by blue arrows in Fig. 3a. This coincides with anomalous behavior of the Hall conductance due to co-existence of electrons and holes when the Lifshitz transition occurs[24] This transition is stretched on the $n_{tot}$ abscissa, since the density of states is large at the singularity. When applying a non-zero displacement, these minima become more pronounced in either direction and no transition occurs. Hence, the displacement field merely improves the resolution of the layer degeneracy removal and the two layers act as one layer irrespective of the size of the displacement field in the range accessible in our experiment. This can be seen in Fig. 3b along the lines of constant $\nu_{tot} = \pm 24$ and $\pm 32$. This behavior is distinct from that at other, lower filling factors, i.e. at significantly lower total densities, where a multitude of displacement field induced transitions can be observed. We assert that this difference stems from the decoupled or weakly coupled nature of the layers at these lower fillings/densities as will be discussed in detail below. The Hall conductance also develops additional features at fillings $\nu_{tot} = \pm 24$ and $\pm 32$, however the observation of a clear transition from $8e^2/h$ steps to $4e^2/h$ steps is hampered by the overall sign reversal of the Hall conductance due to the conversion of the charge carriers from holes to electrons or vice versa as the chemical potential crosses the van Hove singularity energy.

Figure 4 shows a color map of the longitudinal resistance as a function of total filling factor and displacement field as well as single line traces of both the longitudinal resistance and Hall conductance for the low density/low filling regime when the layers are decoupled or weakly coupled only. In the absence of a displacement field, i.e. equal densities in both layers, quantum Hall behavior is only anticipated at even total filling when both layers simultaneously condense in the same quantum Hall state, for instance $\nu_{tot} = \pm 2$ for equal filling 1 of both layers. At non-zero displacement field and fixed total filling, charges are redistributed among the layers. Each layer may then condense in a different quantum Hall state, which would also produce overall incompressible behavior with vanishing resistance. Examples of this can be seen in Fig. 4a and c. At $\nu_{tot} = 2$, the experiment in panel a matches expectations. For $D = 0$, the QHE develops, since both layers take on filling 1. As the displacement field is changed, the top or bottom layer accumulates more charges until it reaches filling 2, while the other layer is emptied resulting in reentrant behavior. This in principle can continue further with hole fillings for one layer at even larger $D$ resulting in additional transitions. The schematic on the right in panel a of Fig. 4 highlights the integer filling of the top and bottom layer at fixed total integer filling for the observed QH-minima in the experimental data on the left. A similar sequence of transitions between QH states also occurs at $\nu_{tot} = 1$ (Fig. 4a) and can be understood in this picture of decoupled layers. However, contrary to expectation, also for $D = 0$ at $\nu_{tot} = 1$, the bilayer system turns incompressible (green dot in the schematic in panel a) as seen in the single line traces of $R_{xx}$ and $\sigma_{xy}$ recorded at 10 T in Fig. 4b as well as at fixed $\nu_{tot} = 1$ for different $B$-fields in Fig. 4c (highlighted by the star). In large twist angle bilayer graphene, only even integer QH states appear for $D = 0$. Odd integer QH physics only emerges for $D \neq 0$.[26] This unexpected $\nu_{tot} = 1$ state at $D = 0$ in our sample is well separated from the QH behavior due to condensation of both layers into their own quantum Hall state by resistance peaks that appear on either side when moving away from $D = 0$. The resistance peaks are followed by reentrant quantum Hall behavior because the top and bottom layer approach fillings $\nu_{top} = 1$ and $\nu_{bot} = 0$ or vice versa. All QH states including the $D = 0$ state and the transitions become more pronounced with increasing $B$-field (Fig. 4c). We have focused here on electron occupation only. Hole transport is discussed in the supporting information, but it is of lesser quality.

In principle, a $\nu_{tot} = 1$ QH state with $D/\varepsilon_0$ close to 0 mV/nm may arise either because of interlayer Coulomb interaction driven Bose Einstein condensation or because a genuine fractional quantum Hall state forms in each of the layers separately, such as for instance at $\nu_{top} = 1/3$ and $\nu_{bot} = 2/3$ or $\nu_{top} = \nu_{bot} = 1/2$. This fractional quantum Hall scenario to produce a $\nu_{tot} = 1$ for $D = 0$ can however be discarded. Our device shows no hint of fractional QH behavior even at 14 T and 30 mK. A half filled fractional quantum Hall state in an isolated graphene monolayer has only been observed so far when the valley degeneracy is lifted as a result of symmetry breaking by a moiré superlattice potential imposed by an adjacent hBN film or at magnetic fields exceeding 30 T.[30] Neither conditions however apply here. Hence, we conclude that the interlayer Coulomb interaction plays a crucial role in reestablishing interlayer coherence in the twisted bilayer. Without interactions, it is absent in essence because of the twist induced moment mismatch of the Dirac cones of the two layers. This is manifested in the absence of odd integer quantum Hall states and odd numerator fractional quantum Hall states in bilayers with large twist angle when $D/\varepsilon_0 = 0$ mV/nm.[27] Our experimental data demonstrate that interlayer coherence and the accompanied exciton condensation can be restored at balanced densities at lower twist angle, resulting in $\nu_{tot} = 1$ quantum Hall behavior. Note that with increasing density or field, i.e. as the chemical potential approaches the van Hove singularity, the states of both layers start to hybridize and the system's behavior should return to that of a single layer. With increasing total density and fields above 12 T, the $\nu_{tot} = 1$ QH state no longer shows clear reentrant behavior when moving away from $D = 0$. The flanking resistance peaks diminish in strength (see Fig. S2 of the supporting information). In this regime, interlayer coherence no longer needs to be invoked to account for the observed physics.

The quantum Hall effect also appears at $\nu_{tot} = 3$ for balanced layer densities. The emergence of this ground state and its evolution at lower fields is illustrated in Fig. 5. The longitudinal resistivity around $\nu_{tot} = 3$ filling as a function of $D/\varepsilon_0$ is plotted as a color map in panel a for different magnetic field values (3 - 6 T). At low fields, this $\nu_{tot} = 3$ QH state centered around $D = 0$ undergoes again a transition heralded by longitudinal resistance peaks with ascending magnitude of the displacement field followed by reentrant quantum Hall behavior because the two layers condense simultaneously in two different quantum Hall states (here filling 1 or 2 and vice versa for the other layer). At higher field, such transition peaks vanish, completely analogous to the behavior at $\nu_{tot} = 1$. This can be seen in the data recorded at 12 T in Fig. S2 of the supporting information. The fields at which the layer coherent state associated with exciton condensation disappears is much lower (about three times) for $\nu_{tot} = 3$ than for $\nu_{tot} = 1$. This however should not come as a surprise as this corresponds to about the same total density where single layer behavior is recovered.

In summary, the peculiar dispersion of twisted bilayer graphene consisting of two displaced Dirac cones, that are each composed of states belonging to one of the constituent layers and that hybridize at higher energy due to magnetic breakdown, offers the unique opportunity to modify the effective interlayer coupling strength or tunneling from very weak to strong simply via the mechanism of magnetic breakdown when tuning the density. The layer degree of freedom is effectively removed as the chemical potential approaches the van Hove singularity. These properties have been exploited here to investigate the appearance and evolution of odd integer quantum Hall states with layer tunneling strength. In the regime of low density when the chemical potential is far away from the van Hove singularity and the layers are weakly coupled, the odd integer quantum Hall effect is observed and attributed to interlayer coherence and the formation of a Bose-Einstein condensate of excitons formed by holes and electrons in half filled Landau levels of the two layers. As the chemical potential is raised and the states of both layers hybridize, conventional single layer quantum Hall physics is restored instead.

## ASSOCIATED CONTENT

**Supporting Information**

Sample fabrication and gating details, Top- and back-gate voltage dependence of the longitudinal resistance, Weakening of the reentrant quantum Hall behavior at $\nu_{tot} = 1$, Effect of the graphene electrode with asymmetric gate tuning, Magnetic breakdown in a bilayer with a 2° twist angle.

## AUTHOR INFORMATION

**Corresponding Author**

**Jurgen. H. Smet** - Max-Planck-Institut für Festköperforschung, 70569 Stuttgart, Germany; Email: j.smet@fkf.mpg.de

**Authors**

**Youngwook Kim** - Max-Planck-Institut für Festköperforschung, 70569 Stuttgart, Germany; Department of Emerging Materials Science, DGIST, 42988 Daegu, Korea

**Pilkyung Moon** - Arts and Sciences, NYU Shanghai, Shanghai 200122, China and NYU-ECNU Institute of Physics at NYU Shanghai, Shanghai 200062, China

**Kenji Watanabe** - Research Center for Functional Materials, National Institute for Materials Science, Tsukuba 305-0044, Japan

**Takashi Taniguchi** - International Center for Materials Nanoarchitectonics, National Institute for Materials Science, Tsukuba 305-0044, Japan

**Notes**

The authors declare no competing financial interest.

**ACKNOWLEDGEMENTS**

We thank K. von Klitzing and D. Zhang for fruitful discussions and P. Herlinger, J. Mürter, Y. Stuhlhofer, S. Göres, and M. Hagel for assistance with sample preparation. This work has been supported by the EU Graphene Flagship (Core 3) and the DFG Priority Program SPP 2244. The work at DGIST is supported by the Basic Science Research Program NRF-2020R1C1C1006914 through the National Research Foundation of Korea (NRF) and also by the DGIST R&D program (20-CoE-NT-01) of the Korean Ministry of Science and ICT. The growth of hexagonal boron nitride crystals was sponsored by the Elemental Strategy Initiative conducted by MEXT, Japan, Grant Number JPMXP0112101001, JSPS KAKENHI Grant Number JP20H00354 and the CREST (JPMJCR15F3), JST. P.M. was supported by the National Science Foundation of China (Grant No. 12074260), the Science and Technology Commission of Shanghai Municipality (Grant No. 19ZR1436400), and the NYU-ECNU Institute of Physics at NYU Shanghai.

**Figure 1**

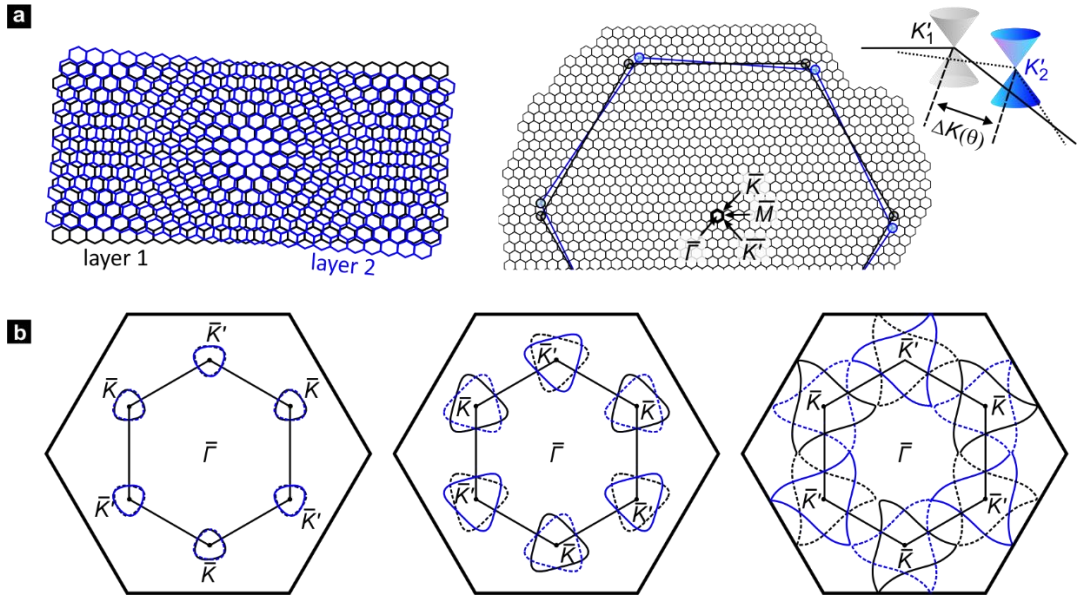

Fig 1. Band structure of twisted bilayer graphene. (a) Real space lattice structure of two graphene layers twisted by an angle θ (left, black – reference layer, blue – twisted layer). The twist in real space causes a displacement of the Dirac cones of the two layers in reciprocal space (right) by an amount determined by the twist angle θ. The black and blue large hexagons indicate the first Brillouin zone of the two layers. The energy where both Dirac cones would intersect increases with increasing twist angle. Also shown is a schematic of the reduced Brillouin zone and the notation used for the symmetry points. (b) For a twisted bilayer, there are two bands. Each of them generate closed isoenergetic contours around the $\bar{K}$ and $\bar{K}'$ symmetry points below the van Hove singularity. The three panels correspond to three different energies. The Fermi contours associated with the first band are plotted with solid lines. Fermi contours that stem from band 2 are plotted with dotted lines. The coloring of the Fermi surfaces indicates whether the states originate from the Dirac cone of the first (black) or second layer (blue).

**Figure 2**

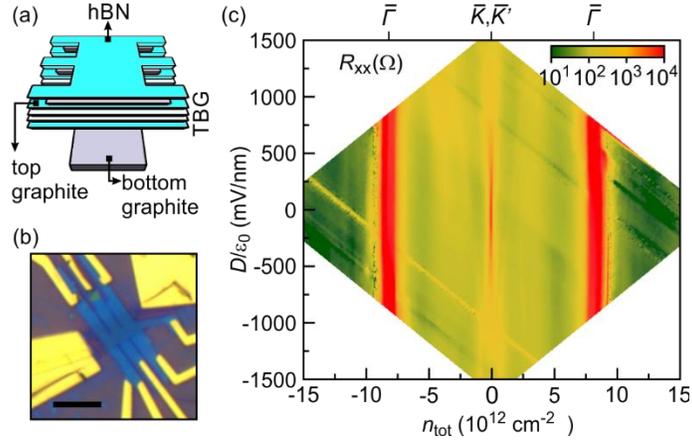

FIG. 2. (a) Schematic of the twisted bilayer graphene device consisting of a graphitic bottom gate, a hBN dielectric layer, the twisted bilayer, a hBN dielectric layer, a graphitic top gate and finally a hBN cap layer. (b) Optical microscope image of the device. The scale bar corresponds to 5 μm. (c) Color map of the longitudinal resistance, $R_{xx}$, in the plane spanned by the carrier density and the displacement field for zero magnetic field. The narrow red region near zero density corresponds to the conventional resistance peak at charge neutrality when the chemical potential crosses the Dirac point at the $\bar{K}$ and $\bar{K}'$-symmetry point. Secondary charge neutrality peaks also appear as red regions near $n_{tot} = +/- 8 \times 10^{12}\,\text{cm}^{-2}$ and signal that the miniband formed due to moiré superlattice potential induced zone folding is either completely emptied or filled at the $\bar{\Gamma}$-point.

**Figure 3**

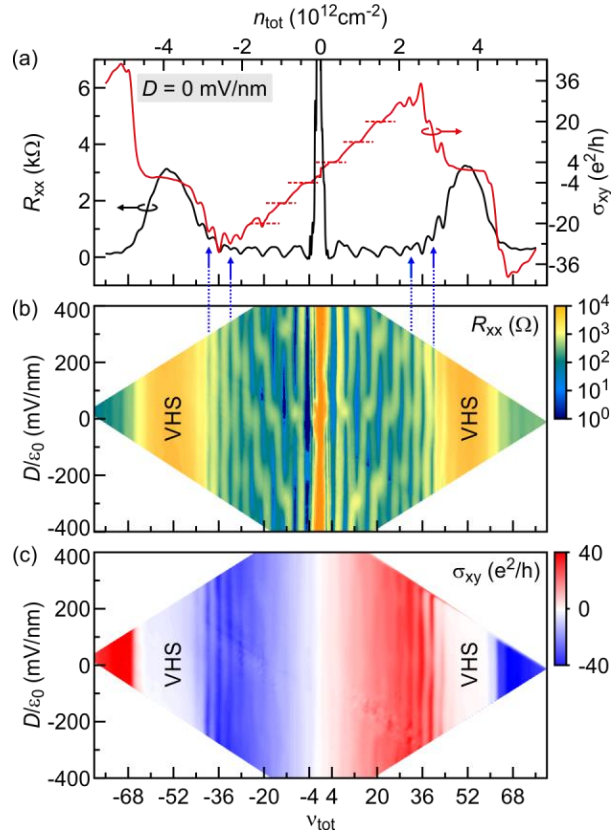

FIG. 3. (a) Longitudinal resistance, $R_{xx}$, and Hall conductance, $\sigma_{xy}$, as a function of the total filling factor or density at a fixed magnetic field of 3 T and in the absence of a displacement field. Red dashed lines highlight the $8e^2/h$ steps in the Hall conductance, for instance at $\nu_{tot} = \pm 4, \pm 12, \pm 20$. Blue arrows and dotted lines point to incipient quantum Hall behavior when the chemical potential approaches the van Hove singularity and the states of both layers hybridize so the system effectively behaves as a single layer. (b) Color map of $R_{xx}$ in the ($\nu_{tot}, D/\varepsilon_0$)-plane. The magnetic field is fixed at 3 T. (c) Same, but for $\sigma_{xy}$.

**Figure 4**

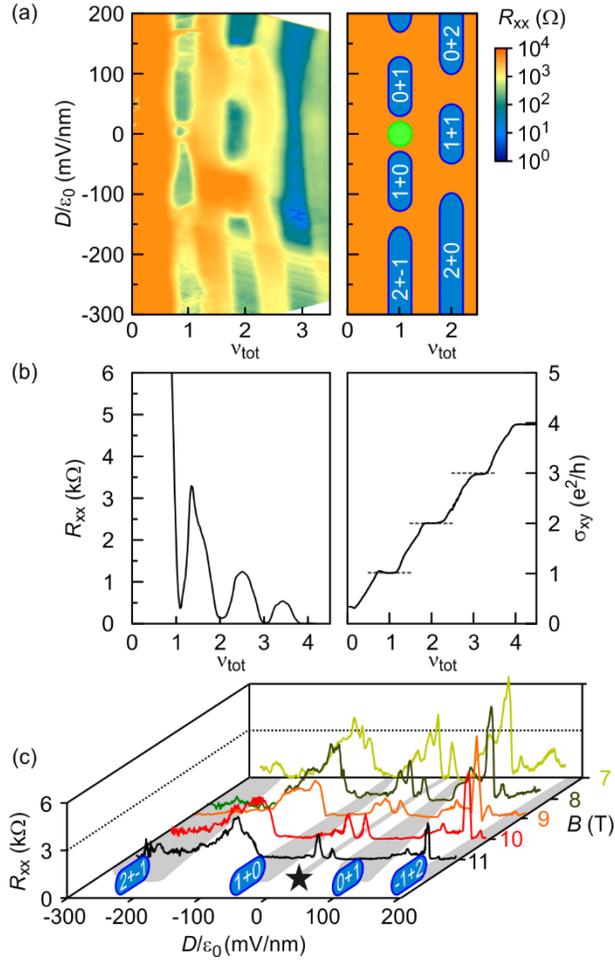

FIG. 4. (a) Color map of $R_{xx}$ (left) as a function of the displacement field and the total filling factor for $B = 10$ T. The schematic on the right indicates regions where quantum Hall behavior is observed due the simultaneous formation of an integer quantum Hall state in each of the layers. The integer numbers correspond to the filling factor of the top ($\nu_{top}$) and bottom layer ($\nu_{bot}$). (b) (Left) Longitudinal resistance, $R_{xx}$, and (Right) Hall conductivity, $\sigma_{xy}$, as a function of $\nu_{tot}$ at 10 T when $D/\varepsilon_0 = 0$ mV/nm. Grey dashed lines mark symmetry broken states. (cS) Traces of the longitudinal resistance as a function of $D/\varepsilon_0$ for different fixed magnetic fields (different colors). All data were recorded at $T = 1.3$ K.

**Figure 5**

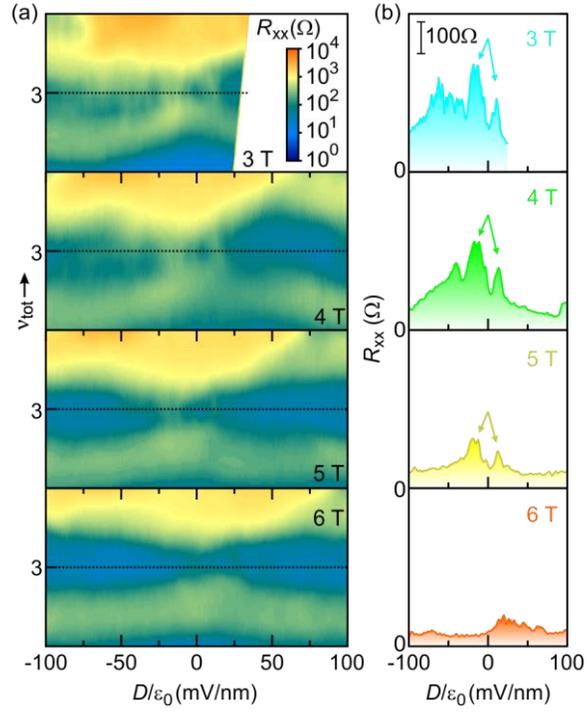

FIG. 5. (a) Color map of $R_{xx}$ near the $\nu_{tot} = 3$ state at magnetic fields from 3 to 6 T (1 T steps). Black dotted lines indicate $\nu_{tot} = 3$. The range of total filling factor covered is 2.5 to 3.5. (b) Line-cuts at fixed $\nu_{tot} = 3$ along the black dotted lines in (a). All windows in (b) have the same y-axis. The vertical bar corresponds to 100 Ω. Arrows mark the resistance peaks signaling the transition from the interlayer coherent ground state to a ground state involving incompressible behavior in each layer separately. These transitions are distinguishable up to about 5 T.

# Supporting Information for "Odd integer quantum Hall states with interlayer coherence in twisted bilayer graphene"

Youngwook Kim[1,2], Pilkyung Moon[3], Kenji Watanabe[4], Takashi Taniguchi[5], and Jurgen H. Smet[1*]

[1]*Max-Planck-Institut für Festköperforschung, 70569 Stuttgart, Germany*

[2]*Department of Emerging Materials Science, DGIST, 42988 Daegu, Korea*

[3]*Arts and Sciences, NYU Shanghai, Shanghai 200122, China and NYU-ECNU Institute of Physics at NYU Shanghai, Shanghai 200062, China*

[4]*Research Center for Functional Materials, National Institute for Materials Science, Tsukuba 305-0044, Japan*

[5]*International Center for Materials Nanoarchitectonics, National Institute for Materials Science, Tsukuba 305-0044, Japan*

## S1. Sample fabrication and gating details

The van der Waals heterostructure consists of a stack of seven layers fabricated with the dry stamp pick-up transfer method[S1] supported by a thermally oxidized Si substrate: graphitic bottom gate, hBN, twisted bilayer, hBN, graphitic top gate and hBN cap layer. The hBN layers and the graphite layers have a thickness between 5 and 10 nm. The twisted bilayer is identical to the one that has been studied in Ref. S2. However, the device reported at that time suffered from a short circuit between the front- and back-gate. This short-circuit was successfully removed by performing an additional e-beam lithography step and an etching step. The availability of a top and bottom gate allows for the application of a displacement electric field $D$ between the two graphene layers. In the color rendition of the longitudinal resistance recorded across the plane spanned by the total density $n_{tot}$ and the displacement electric field $D$ plotted in Fig. 1c of the main text, $n_{tot}$ follows from $(C_{TG}V_{TG}+C_{BG}V_{BG})/e - n_0$. Here, $V_{TG}$ and $V_{BG}$ are the top and bottom gate voltages, and $C_{TG}$ and $C_{BG}$ are the capacitances per unit area. The elementary charge is denoted as $e$. The residual charge when both gate voltages are set to zero is referred to as $n_0$. The above capacitances were determined experimentally from the gate voltage dependencies of the Shubnikov-de Haas oscillation frequencies. The displacement electric field is calculated from the expression $D/\varepsilon_0 = (C_{TG}V_{TG} - C_{BG}V_{BG})/2 - D_0/\varepsilon_0$. Here, $\varepsilon_0$ is the permittivity of free space and $D_0$ is the residual electric field originating from residual charges in both layers.

## S2. Top- and back-gate voltage dependence of the longitudinal resistance

Panel a of Fig. S1 illustrates the dependence of the four terminal longitudinal resistance on the back-gate voltage for fixed zero top-gate voltage, whereas panel b shows data as a function of the top-gate voltage for zero back-gate voltage. These measurements were performed at 1.3 K. As mentioned in the main text, three main features are visible. They occur when the chemical potential aligns with the Dirac point at the $\bar{K}$-symmetry point of the Brillouin zone (middle) and when the mini-band around zero energy is completely occupied (right feature) or emptied (left feature). The latter two cases occur near the $\bar{\Gamma}$-point and also result in a resistance peak as it resembles charge neutral conditions. The density change required to fully occupy the miniband provides an estimate of the twist angle, which equals approximately 2°.

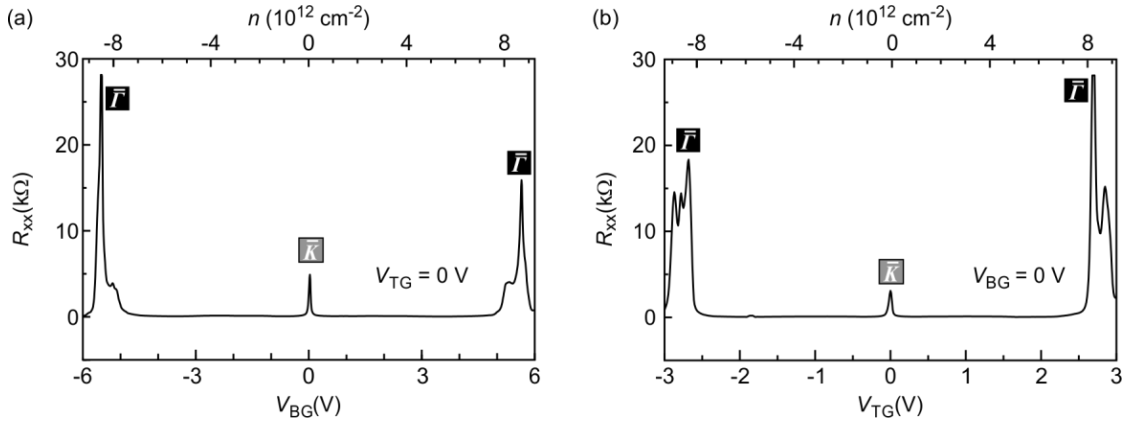

**Fig. S1 | Dependence of the longitudinal resistance with top-gate voltage and back-gate voltage in the absence of a magnetic field.** (a) The resistance curve for a varying back gate (density) with a zero bias top gate. (b) As in (a), but when sweeping the top gate voltage at a fixed back gate voltage of 0 V instead.

## S3. Weakening of the reentrant quantum Hall behavior at $\nu_{tot} = 1$

With increasing magnetic field/density, the resistance features that appear as a result of reentrant quantum Hall behavior at $\nu_{tot} = 1$ when moving away from zero displacement field eventually weaken and vanish as can be seen in Fig. S2. This is similar to what is discussed in the main text for $\nu_{tot} = 3$. The arrows in Fig. S2 mark the resistance peaks separating the incompressible ground state due to Bose-Einstein condensation with interlayer coherence from quantum Hall behavior at non-zero displacement due the condensation of both layers separately in different integer quantum Hall states. The reentrant resistance features remain clearly visible up to 11 T, but weaken significantly at higher fields.

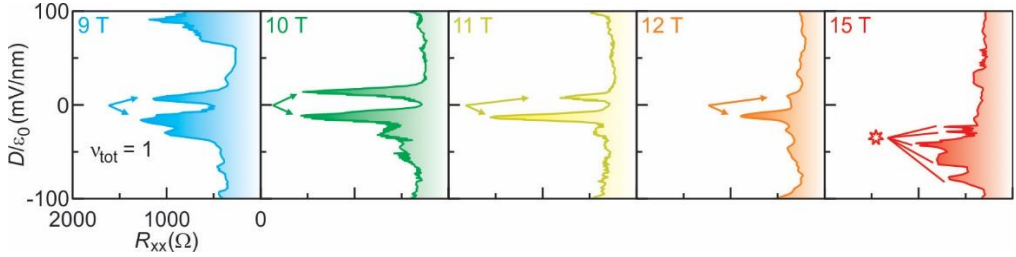

**Figure S2** | Longitudinal resistance at $\nu_{tot} = 1$ as a function of $D/\varepsilon_0$ for different values of the magnetic field (from left to right): 9, 10, 11, 12, and 15 T. All windows have the same abscissa covering a longitudinal resistance from 0 to 2 k$\Omega$. The red star with arrows in the panel for 15 T marks resistance features that result from low conductivity of the graphene contact legs as explained in section S3. This artefact is most pronounced when the areas not covered by the bottom gate enter the true insulating state at $\nu_{tot} = 0$.[S3,S4] It causes strong fluctuations of the resistance.

## S4. Effect of graphene electrode regime with asymmetry gate tuning

The device that we focused on in this manuscript has two graphitic gates, however these gates do not have the same size. More specifically, the top graphite gate covers a larger area of the twisted graphene bilayer than the bottom graphite gate. This is schematically illustrated in panel a of Fig. S3. As a result, part of the graphene bilayer cannot be controlled by the bottom gate. This is the case here in areas of the bilayer that serve as contact legs to the rectangular Hall bar shape. These areas may turn highly resistive, for instance near the charge neutrality point, when the miniband is completely full or empty or when an incompressible ground state such as the $\nu = 0$ state forms in these areas. In Fig. 1(a) such incidents where contacts become poor are visible as parallel diagonal streaks of high resistance. The effect can also be seen in Fig 3(a) and Fig S3(b).

This issue has been discussed previously in the literature studying the fractional quantum Hall effect in bilayer graphene.[S5] In that report, the top and bottom gate were patterned with electron beam lithography to ensure an identical shape of both the top and bottom gate. The graphene based contact legs inevitably still have some area not covered by the gate. However, it is possible to use the doped Si substrate back gate to modify the filling and make sure these areas are conducting well. While in the device presented here, a Si back gate is also available, it is unfortunately not possible to ensure that the entire contact leg area is conducting well, since in regions where the two graphitic gates do not overlap, different densities/fillings are imposed. Fig. S3(b) is an extended version of the color map shown in Fig 3 of the main text covering both electron and hole densities. Although the hole side shows similar quantum Hall states and reentrant quantum Hall transitions, highly resistivity contact leg areas produce diagonal features near zero displacement field. As a result, the quantum Hall features and their transitions are less clear on the hole side than on the electron side.

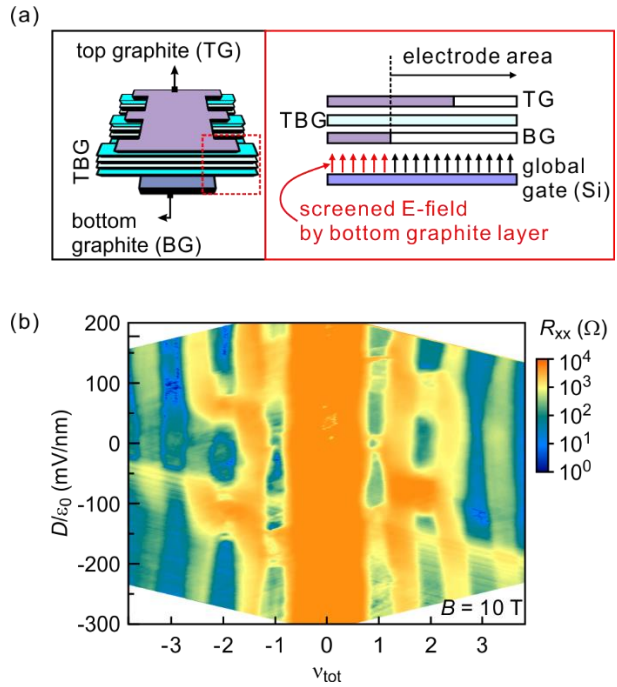

**Figure S3** | (a) Left: Schematic of the device. Right: Schematic illustrating the different overlap of the top (TG) and bottom (BG) graphitic gates with the twisted bilayer. The area of the twisted bilayer not covered by the bottom nor the top gate can always be made conducting by applying a suitable voltage to the second, global back gate formed by the doped silicon substrate. Here, typically $|V_{global}| > 50$ V. The active device area itself is not influenced by this global gate voltage, since the electric field is screened by the bottom gate. (b) Color map of the longitudinal resistance as a function of $\nu_{tot}$ and displacement field. This figure is the same as Fig 3(a) in the main text, but covers a larger range of total filling factor including hole doping.

## S5. Magnetic breakdown in a bilayer with a 2° twist angle

At first sight, it may be surprising that the interlayer coherent states is already suppressed at relatively low density and field (10 T). Usually, magnetic breakdown is prominent for Landau levels close to the van Hove singularity. However, in a bilayer with a 2° twist angle the van Hove singularity already occurs at ±0.025 eV and is much closer to the charge neutrality point (i.e., Dirac point energy, E=0) than the band edges ($\approx$±0.1 eV). Magnetic breakdown starts well before the electron density is such that the chemical potential reaches the van Hove singularity. and a small magnetic breakdown probability may hamper the formation of the BEC state. Figure S4 shows the energy spectrum for the bilayer with a 2° twist angle of 2° plotted against the normalized magnetic field $B/B_0$ (abscissa) and the energy in eV (ordinate). We refer to Ref. S6 for more details about this calculation. For this system $B_0$ equals 96.2 T, and hence an externally applied field of 10 T corresponds to a normalized field of about 0.10. The first Landau level already exhibits a finite "band width" below $B/B_0$=0.10. This can be interpreted as evidence for magnetic breakdown between the electron orbits from the two layers (see also Fig. 6 in Ref. S6). As a result the decoupled nature of the layers is lost quickly and the system can no longer maintain the BEC state.

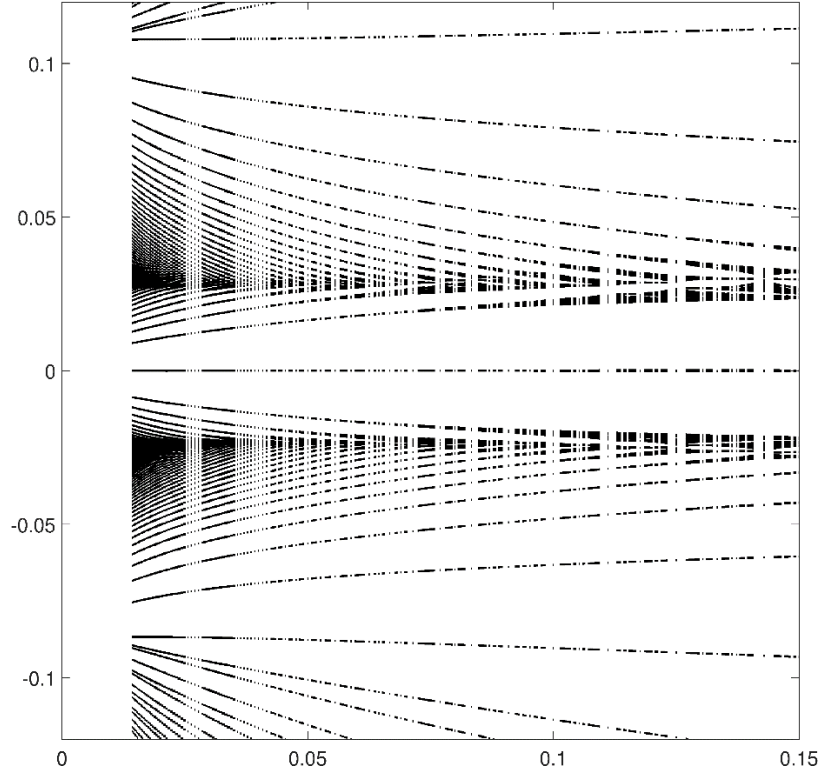

**Figure S4|** Energy spectrum of twisted bilayer graphene with a twist angle of 2° plotted against the normalized magnetic field $B/B_0$ (lateral axis, $B_0$=96.2 T) and energy [eV] (vertical axis).

## References for Supporting Information